# Experimental Polarization Control of Thomson Scattering X-ray Source

Hongze Zhang,* Yingchao Du, Chuanxiang Tang, Lixin Yan and Wenhui Huang
Department of Engineering Phiysics, Tsinghua University, Beijing 100084, China

**Abstract** : Thomson scattering of intense laser pulses from relativistic electrons allows us to generate high-brightness and tunable-polarization X/γ-ray pulses. In this paper we demonstrate the polarization control of the Thomson scattering source experimentally. We control the incident laser polarization by rotating a quarter-wave plate, thus controlling X/γ-ray polarization. In order to measure the polarization, we use Compton scattering method. Meanwhile, stokes parameters of X/γ-ray whose energy varies between tens of keV and MeV are simulated. The simulation results show that with the increasing of X-ray Energy, X-ray polarization is a constant value in a small cone of motivation. According to modulation curves analysed from experiment results, we can get the conclusion that the polarization of Thomson scattering source is tunable and controllable.

Polarized X/γ-ray has been studied and widely used in various scientific field. Polarization of X/γ-ray can provide unique information addition to X-ray imaging and analysis of spectroscopy for researchers. In astrophysics, polarimetric obesvations of neutron stars provide the information of the intensity and geometry of the magnetic field [1-4]. In material science and biology, porlarized X-rays can enhance the sensitivity of X-ray fluorescence analysis [5]. In nuclear physics, polarized γ-rays plays an important role in studying nuclear property. We can study the structure of nuclear by nuclear resonance fluorescence with polarized X/γ-ray (NFR)[6-9]. And the polarized state of γ-rays are also important for the measurements of the parity of the nuclear states[10], the investigation of giant resonances of nuclei and the scattering reactions between photons and nuclei [11,12].

Compton scattering is the elastic scattering of a photon from a free electron, for the low energy electron (~MeV), it is also called as Thomson scattering [13-15]. Compton (Thomson) scattering X-ray source has been studied and developed for decades [16-20]. Comparing to the mechanism of other radiation sources, it can produce unltra-short, energy continuously tunable, high brightness, well-collimate and high polarized X-ray beams by laser photons scattering from free relativistic electrons [21-24]. Because of the advantages in X-ray application, Thomson scattering X-ray source is utility in material, medical and biological areas [24-28]. In our experiment, we change the polarization of X-ray by adjusting the polarization of laser

*zhz16@mails.tsinghua.edu.cn

beams since the polarization of laser is directly transferred to the scattered photons.

In classical shemes, while the intensity of electric field for accelerating electron is weak enough, the electromagnetic field generated by motivating electron can be cast in the form as [29]

$$\vec{E} \sim \frac{1}{r} \frac{\vec{e}_r \times (\vec{e}_r \times \vec{E}_L)}{\left(1 - \frac{\vec{\beta} \cdot \vec{e}_r}{c}\right)^3}$$

where $\vec{e}_r$ is the unit vector from the electron to the observer, r is the distance between the observer and electron, $\vec{\beta}$ is the normalized velocity of the electron, c is the speed of light and $\vec{E}_L$ is the intensity of the electric field of incident laser beams. Polarization of light describes the property of transportation of electromagnetic field in space. In the formula of electron radiation, $\vec{E}$ is proportional to $\vec{E}_L$. So Thomson scattering X-ray sources' polarization are directly determined by incident laser beams' polarized states.

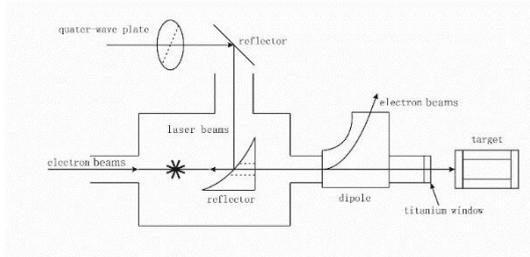

Figure 1 schematic of the experiment

In order to control the polarization of Thomson Scattering X-ray sources precisely, we verify the relationship of polarization between incident laser beams and X-ray beams. We carry out the X-ray polarization control and measurement experiment on Tsinghua Thomson scattering X-ray source (TTX) platform. TTX is set up with a linac system and a femtosecond laser system. The linac system consists of a S-band photo-cathode RF gun, a magnet compressor and two x-band harmonic structures to generate high brightness electron pulse. The laser system can generate 266-nm ultraviolet pulse for the photocathode and 800-nm infrared pulse for the scattering interaction. The energy of X-ray photons is 50-keV and the flux is about $10^7$-s$^{-1}$[30]. In our experiment, laser photons track through the quarter-wave plate and have a head-on interaction with the high quality electron beams in the vacuum interaction room. The polarization of scatted photons is determined by the incident laser beams, which are controlled by the quarter-wave plate precisely.

The polarization property of Thomson scattering scources has been study by several groups few years ago [31-34]. They distinguish different polarized states and helicity of circle polarization of X/γ by observing scattered photons distribution, energy spectrum analysis in space fixed points and transmission method. However, in our experiment, we use the Compton scattering method, a kind of polarization-sensitive process which is more accurate than before, to measure the polarization of X-ray beams. According to the Klein-Nishina formula [35], for linear incident X/γ-ray photon, the cross section of photon scattering from free electrons is:

$$\frac{d\sigma}{d\Omega} = \frac{r_0^2}{2}(\frac{E'}{E})^2(\frac{E'}{E} + \frac{E}{E'} - 2\sin^2\theta\cos^2\Phi)$$

where $r_0$ is classical electron radius, θ and φ are azimuthal angles, E and E' are the incident photons and scattered photons energy. The azimuthal distribution of the scattered photons are strongly depended on the X-ray polarization.

A target, made from polyethylene, is placed after the titanium window of the beam pipe. The size of the cylinder target is 5-cm in height and 0.75-cm in radius. X-ray pulses irradiate on the end of the cylinder and generate scattered photons. We use an image plate wrapping around the cylinder to record scattered photons(figure 2). Meanwhile, we use two thin aluminum rings locked to both ends of the polyethylene cylinder to support the image plate. The curved image plate is 2.5-cm in radius.

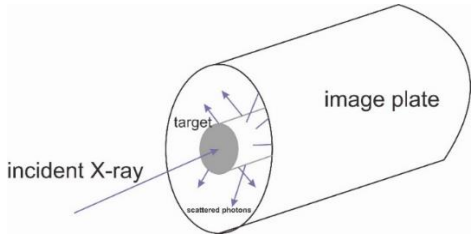

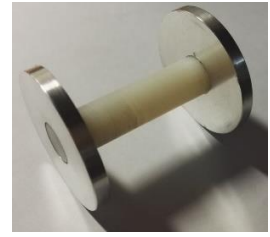

Figure 2. Schematic and real picture of taget.

Figure 3 shows the experiment results recorded by image plate and the simulation results done with Geant4.

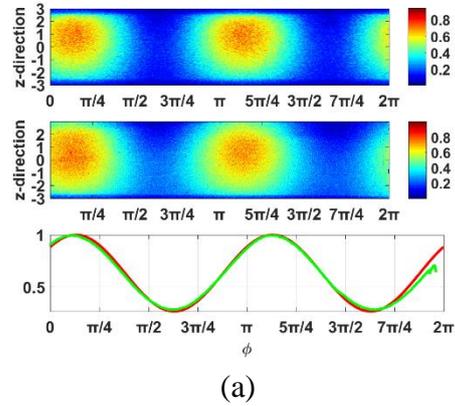

(a)

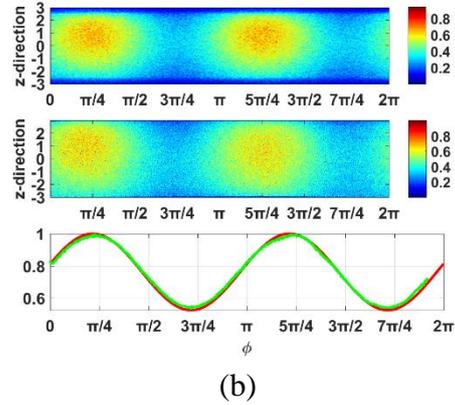

(b)

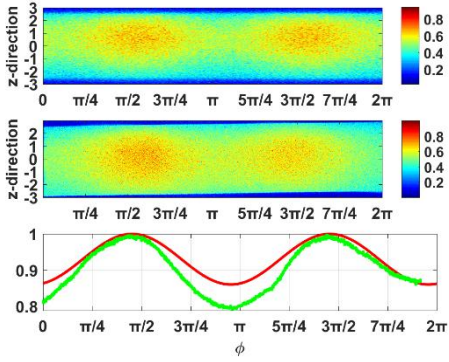
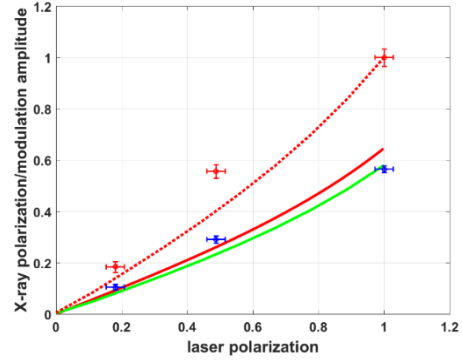

(c)

Figure 3. (a) Linear polarized incident laser beams. First line: Simulation result of the scattered photons recorded by the image plate. Second line: experiment result of the scattered photons recorded by the image plate. Third line: Calculated results from the simulation (red) and experiment (green) results. (b) Ellipse polarize incident laser beams. (c) Circle polarized incident laser beams.

Figure 4. Red line: the theory result of modulation amplitude. Green line: the simulation result of modulation amplitude. Red dot line: the simulation result of polarization. Red spots: experiment results of polarization. Blue spots: experiment results of modulation amplitude.

In figure 3, for each polarization result, we sum the photons recorded by the image plate in the z-direction and get modulation curves. According to the modulation curves, we calculate the modulation amplitude and polarization of different polarized X-ray pulses. The modulation curves' amplitudes vary from the maximum value for linear polarization to the minimum value for circle polarization. The reasons, affecting the accuracy of the results, include the jitter of the electron pulse and laser pulse. Also, we assume that the X-ray bunches are parallel after tracking through the filter. But there is still a small divergence for each X-ray bunch in the forward direction.

Simulated result (green line) is smaller than the theory result (red line). Because when we calculate the theory result, we only consider primary scattering process and ignore the background. The variation trend of the experiment result fits the simulation results. However, there is still one point doesn't fit the simulation results well. Because the simulation process is done under an ideal condition. The polarization of X-ray irradiated on the target is supposed to be the same. However, X-rays generated by Thomson scattering sources have a small divergence in forward direction. It results in the polarization of X-ray irradiated on the target varies between a small region. We can reduce the target's cross section radius to improve the accuracy of the experiment.

The X-ray energy is about 50-keV in

our experiment and the X-ray polarization is nearly a constant value. In order to verify the polarizaiton relationship between incident laser pulses, electron beams in high energy section, we calculate stokes parameters of X-rays under different electron beams energy. Limited by the experiment condition, we simulate the interaction process with Cain program. Stokes parameter $S_2$ represents the circle polarization and $S_3$ represents the linear polarization. $S_2 = 1$ or $S_2 = -1$ represents clockwise and counterclockwise. And $S_1(S_3) = 1$ or $S_1(S_3) = -1$ represents two polarized directions which are orthogonal to each other. Complete polarized states have $S^2 = 1 (S^2 = S_1^2 + S_2^2 + S_3^2)$, but mixed states have $S^2 < 1$. Changing energies of electron beams in Cain, the average value of $S_2$ and $S_3$ in a small cone ($\theta = 1/5\gamma$) under different X-ray energys are shown in figure 5.

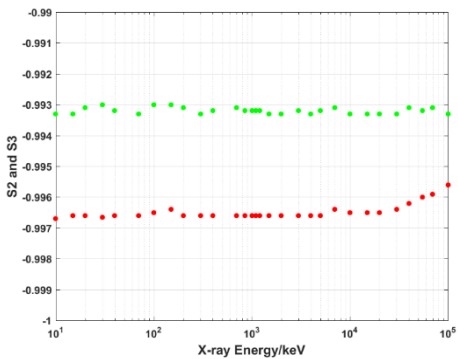

Figure 5. Stokes parameters. Green line: average value of $S_2$ under different X-ray energy. Red line: average value of $S_3$ under different X-ray energy.

In figure 5, Stokes parameters are almost invariant with X-ray energy increasing in a constant small cone($\theta = 1/5\gamma$). This means energies of electron beams doesn't affect the polarization of X-ray with a wide range of energy in a small cone. In order to measure the high energy X-ray polarization, a smaller cross section target is useful. But with high energy X-ray, about 10-Mev or more, the effect of pair production will affect the accuracy of the measurement.

**Conclusion**: Thomson Scattering Source is an important way to generate polarized X/γ -ray pulses which are tunable and controllable. Polarized X/γ-ray pulses are important and useful in various scientific area. Our experiment is the first time that try to precisely measure the polarization of Thomson Scattering Sources and verify the relationship of polarization between the incident laser beams and scattered X-ray beams. We use Compton scattering method, applicable in the energy range keV ~ MeV, to measure the polarization of X-ray beams. The experiment results show that we can produce accuracy polarization γ-ray/X-ray pulses by changing the polarized state of incident laser beams. And X-ray polarization are nearly constant value in a small cone under different X-ray energy.